\documentclass{article}
\usepackage{graphicx} 
\usepackage{subfigure}
\usepackage{bm}
\usepackage{amsmath}
\usepackage{amssymb}
\usepackage{xcolor}
\usepackage{amsthm} 
\usepackage{mathtools}
\usepackage{authblk}
\usepackage{hyperref}
\usepackage{natbib}
\usepackage{amsmath}
\DeclareMathOperator{\tr}{tr} 
\providecommand{\keywords}[1]{{{Keywords---}} #1}
\usepackage{orcidlink}
\title{{Permutation invariant neural network prediction of vacancy formation under deformation and varying chemical environment in FCC high entropy alloys}\footnote{Notice: This manuscript has been coauthored by UT-Battelle, LLC, under Contract No. DE-AC0500OR22725 with the U.S. Department of Energy. The United States Government retains and the publisher, by accepting the article for publication, acknowledges that the United States Government retains a non-exclusive, paid-up, irrevocable, world-wide license to publish or reproduce the published form of this manuscript, or allow others to do so, for the United States Government purposes. The Department of Energy will provide public access to these results of federally sponsored research in accordance with the DOE Public Access Plan (\href{http://energy.gov/downloads/doe-public-access-plan}{http://energy.gov/downloads/doe-public-access-plan}).}}
\author[1]{Tanvir Sohail\orcidlink{0000-0001-7567-6417}}
\author[1*]{Swarnava Ghosh\orcidlink{0000-0003-3800-5264}}
\affil[1]{National Center for Computational Sciences, Oak Ridge National Laboratory, TN 37830}
\affil[*]{Email: ghoshs@ornl.gov}

\begin{document}

\maketitle
\begin{abstract}
Vacancy formation energies govern diffusion, irradiation damage, phase stability, and dynamic failure in high-entropy alloys (HEAs), yet their strong dependence on local chemical environments and mechanical deformation makes atomistic calculations prohibitively expensive for large-scale studies. Here, we develop an atomistically informed permutation invariant machine learning framework for predicting strain dependent vacancy formation energies in FCC HEAs from local atomic environments. The model employs a vacancy-centered representation constructed from objective geometric descriptors together with invariants of the local deformation gradient, enabling the coupled effects of chemical disorder and finite deformation to be learned within a unified framework. Atomistic simulations reveal that volumetric deformation is the dominant factor controlling the average variation in vacancy formation energy, whereas shear deformation has a comparatively minor influence. At the same time, substantial site to site variability persists under identical macroscopic loading, demonstrating that local chemical environments govern the statistical distribution of vacancy energetics beyond species-averaged trends. The proposed framework accurately predicts vacancy formation energies across diverse deformation states while providing orders-of-magnitude faster evaluation than direct atomistic simulations. These results establish an efficient route for incorporating stress-dependent defect energetics into multiscale models of diffusion, irradiation damage, and dynamic failure in chemically complex alloys.
\end{abstract}
\section{Introduction}

Vacancies are among the most fundamental crystal defects and play a central role in determining the mechanical behavior, diffusion kinetics, and microstructural evolution of crystalline materials. Their interactions with other vacancies and solutes give rise to elastic and electronic perturbations that may extend over several unit cells \cite{ghosh2022spectral}. The formation of vacancies and their evolution under applied stress strongly influence a wide range of phenomena, including hydrogen embrittlement through vacancy clustering and void nucleation \cite{lu2005hydrogen}, vacancy-assisted dislocation motion and low-temperature softening \cite{lu2002can}, irradiation-induced hardening via the formation of vacancy clusters and prismatic loops \cite{gavini2007vacancy}, and spall under shock loading through vacancy nucleation and coalescence \cite{ho2007energetics}. Consequently, understanding how mechanical deformation modifies vacancy formation energetics is essential for predicting the performance and reliability of structural materials under extreme loading conditions.

High-entropy alloys (HEAs) comprise multiple principal elements in near-equiatomic proportions that often stabilize single-phase solid solutions over a wide compositional space \cite{yeh:2004,cantor:2004,george:2019:Nature,george:2020,varvenne:2017}. The interplay between configurational entropy and enthalpic interactions suppresses competing intermetallic phases, giving rise to chemically disordered alloys with exceptional mechanical and environmental performance \cite{yeh:2004,george:2019:Nature,george:2020,varvenne:2017,anand2026edge,chabri2026comprehensive}. The most extensively studied single-phase HEA families include the face-centered cubic (FCC) Cantor alloy class, body-centered cubic (BCC) refractory HEAs, and rare-earth HEAs with hexagonal close-packed (HCP) crystal structures, each exhibiting distinct defect energetics and transport characteristics. Their outstanding structural and functional properties have motivated applications in extreme environments, including nuclear energy systems, where defect generation, migration, and clustering govern long-term reliability \cite{radiation,thermoelectric,softmagnetic}. In contrast to many conventional alloys, which can readily undergo stress-induced precipitation \cite{ghosh2020influence,ghosh2021precipitation}, high-entropy alloys (HEAs) often retain remarkably stable solid-solution phases. Their chemically complex local environments give rise to pronounced site-to-site variability in vacancy formation and diffusion energetics \cite{ponga2022,Fani}. Consequently, understanding vacancy energetics in the presence of severe chemical disorder has emerged as a central challenge in the computational design of HEAs.

Vacancies play a central role in governing diffusion, phase transformations, and irradiation response in high-entropy alloys (HEAs). Compared with pure metals and conventional binary alloys, HEAs exhibit significantly higher equilibrium concentrations of vacancies and vacancy clusters, leading to profound changes in their thermodynamic and kinetic behavior \cite{wang2017thermodynamics}. This enhanced vacancy activity has also been demonstrated experimentally, where electropulsing treatment increases vacancy concentrations and diffusion rates in Cr$_{22}$Mn$_{20}$Fe$_{21}$Co$_{18}$Ni$_{19}$, promoting Cr--vacancy clustering and accelerating $\sigma$-phase precipitation \cite{wu2026rapid}. Positron lifetime measurements on the equiatomic CoCrFeMnNi Cantor alloy have further reported a vacancy formation enthalpy of approximately $1.7$ eV and diffusion activation energies comparable to those of conventional FCC metals, providing direct experimental constraints on vacancy energetics in chemically complex alloys \cite{Sugita2020}.

\begin{figure}[ht!]\centering
{\includegraphics[keepaspectratio=true,width=0.75\textwidth]{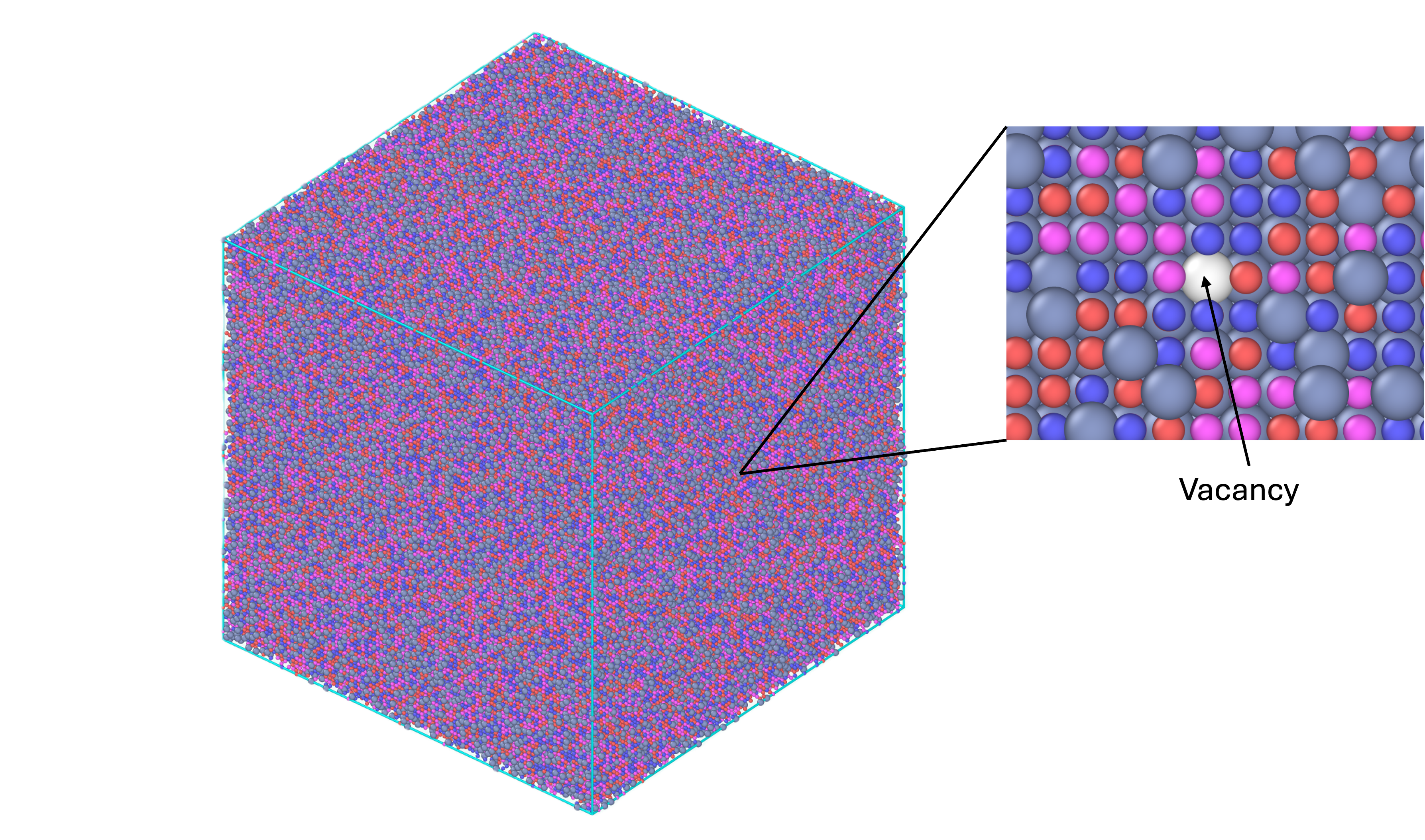}}
{\caption{Atomic supercell of FCC CoCrFeNi high entropy alloy is shown. The inset shows a vacancy in this alloy with its local chemical environment.}\label{Fig:schematic}}
\end{figure}

First-principles calculations have subsequently revealed that vacancy energetics in HEAs are governed predominantly by the local chemical environment rather than the average alloy composition. Figure \ref{Fig:schematic} shows the local chemical environment surrounding a vacancy in a typical FCC high entropy alloy. In equiatomic CoCrFeMnNi, vacancy formation energies are approximately $2.0$ eV on average, whereas migration barriers exhibit substantial variations arising from local lattice distortions and electronic structure effects \cite{Mizuno2019}. Similar local-environment dependence has been reported in CuNiCo and CuNiCoFe alloys, where both alloy composition and temperature significantly modify effective vacancy formation energies, with Cu exhibiting the strongest interaction with vacancies \cite{Esfandiarpour2019}. Recent calculations on CoCrFeMnNi have further demonstrated that both vacancy formation and migration energies are broadly distributed because of chemical disorder, resulting in heterogeneous diffusion pathways that cannot be described by a single Arrhenius activation energy \cite{Wang2022}. Such heterogeneous vacancy energetics have also been proposed as a key origin of the exceptional irradiation tolerance of HEAs through their influence on vacancy transport and defect evolution \cite{gao2024defect}.

The local chemical environment also strongly affects the stability and evolution of vacancy clusters. First-principles studies have shown that unstable small vacancy clusters suppress stacking fault tetrahedra formation in CoCrFeMnNi under moderate irradiation, although larger vacancy clusters may become stable under cascade damage \cite{xu2021irradiation}. Likewise, ab initio simulations of CoCrFeNi have revealed broad distributions of vacancy-cluster formation and binding energies that promote defect-cluster dissociation and substantially influence vacancy aggregation, diffusion, and irradiation-induced microstructural evolution \cite{luo2025ab}. Collectively, these studies demonstrate that vacancy energetics in HEAs are intrinsically statistical, reflecting the underlying chemical disorder, and therefore require predictive models capable of resolving local atomic environments rather than relying on average alloy properties.

The complexity of local chemical environments in high-entropy alloys has made machine learning (ML) an increasingly important tool for predicting defect properties that are otherwise prohibitively expensive to compute using first-principles methods. Early studies combined ML with kinetic Monte Carlo simulations to establish composition--structure--property relationships governing hydrogen transport in FeCoNiCrMn alloys, enabling the identification of compositions with reduced hydrogen diffusivity and improved resistance to hydrogen embrittlement \cite{Zhou2022H}. More recently, ML models have been used to learn the local environment dependent potential energy landscape of HEAs, revealing that chemically induced distributions of site energies and migration barriers govern sluggish self-diffusion \cite{Xu2022}.

The application of ML to vacancy energetics has also advanced rapidly. A graph-based framework combining the CHGNet universal interatomic potential with crystal graph convolutional neural networks (CGCNNs) has been developed to predict vacancy formation energies directly from relaxed atomic structures. By incorporating DFT derived electronic descriptors, the model achieved accurate and transferable predictions across chemically complex alloy systems while substantially reducing the need for explicit first-principles calculations \cite{linton2026framework}. In parallel, support vector regression (SVR) has been applied to predict vacancy formation energies in random solid-solution and locally chemically ordered HEAs. By systematically comparing local atomic descriptors, this study identified smooth overlap of atomic positions (SOAP) descriptors as the most effective representation of local chemical environments, highlighting the critical importance of descriptor selection and chemical ordering for accurate prediction of defect energetics \cite{ibrahim2026machine}.

Despite these advances, developing machine-learning models that accurately capture the interplay between local chemical disorder, atomic structure, and defect energetics while remaining computationally efficient and transferable across diverse alloy chemistry remains an important challenge for the accelerated design of high-entropy alloys.

Dynamic loading generates large non-equilibrium populations of vacancies that govern void nucleation, spall failure, and irradiation damage in crystalline materials \cite{ghosh:2019,meyers1994dynamic}. In FCC high-entropy alloys (HEAs), shock-induced defect evolution is further complicated by severe chemical disorder, which promotes heterogeneous plasticity through dislocation interactions, stacking faults, and deformation twinning while simultaneously altering vacancy generation and transport \cite{zhao2023deformation,majeed2025vacancy,peng2022role}. Although recent experimental and atomistic studies have provided valuable insights into shock-induced deformation and spallation in FCC HEAs \cite{zhao2023deformation,majeed2025vacancy,peng2022role,song2023dynamic,jiang2016shock}, the underlying role of stress in modifying vacancy energetics remains largely unexplored. Existing atomistic investigations of vacancy formation have almost exclusively considered stress-free configurations, even though vacancy formation energy is intrinsically coupled to the local stress state. Consequently, a predictive atomistic framework capable of quantifying the combined effects of mechanical loading and local chemical environments on vacancy formation energies is still lacking.

In this work, we present an atomistically informed machine-learning framework for predicting strain- and chemical-environment-dependent vacancy formation energies in FCC high-entropy alloys. By combining atomistic calculations under finite deformation with local atomic environment descriptors, the proposed model quantifies the coupled effects of chemical disorder and mechanical loading on vacancy energetics across diverse atomic configurations. This framework overcomes the limitations of conventional zero-stress defect calculations and provides an efficient route for incorporating stress-dependent vacancy thermodynamics into multiscale models of diffusion, irradiation, and dynamic failure in chemically complex alloys.

The remainder of this paper is organized as follows. In Section \ref{Sec:results} we first examine the influence of mechanical deformation on the species-averaged vacancy formation energy in FCC CoCrFeNi. We then introduce a permutation invariant neural network for predicting vacancy formation energies across diverse local chemical environments and deformation states, and establish the rotational invariance of the underlying descriptor representation. The predictive performance of the model is subsequently evaluated and applied to quantify the effects of deformation on site-specific vacancy formation energy, entropy, and equilibrium concentration. Finally, we conclude with a discussion of the implications and limitations of the proposed framework, outline directions for future research, and describe the computational methodology.

\section{Results}\label{Sec:results}

\subsection{Effect of deformation on vacancy formation energy}\label{Sec:deformationVFE}
Mechanical deformation modifies both the total energy of the crystal and the chemical potentials of its constituent elements, thereby altering the thermodynamics of vacancy formation. To quantify this effect, vacancy formation energies are evaluated under finite deformation of the crystal. The deformation is described by the deformation gradient tensor $\mathbf{F}$, which maps the atomic position vector $\mathbf{R}_i$ in the reference configuration to the corresponding position vector $\mathbf{r}_i$ in the deformed configuration through $\mathbf{r}_i = \mathbf{F}\mathbf{R}_i$.

The corresponding Green--Lagrange strain tensor is
\begin{equation}
\mathbf{E}=\frac{1}{2}\left(\mathbf{F}^{T}\mathbf{F}-\mathbf{I}\right).
\end{equation}

For a supercell containing $N$ atoms, the vacancy formation energy associated with removing an atom of species $A$ is defined as
\begin{eqnarray}
    E^{vf}_A (\mathbf{F}) = {E}_{A}(N-1;\mathbf{F}) - {E}_{0}(N;\mathbf{F}) + \mu_A(\mathbf{F}) \,\,,
\end{eqnarray}
where $E_A(N-1;\mathbf{F})$ and $E_0(N;\mathbf{F})$ denote the total energies of the defective and perfect supercells under the same deformation, respectively, and $\mu_A(\mathbf{F})$ is the deformation-dependent chemical potential of species $A$.
The chemical potentials are obtained self-consistently from substitution calculations. Specifically, the energy difference associated with replacing an atom of species $A$ by species $B$ is related to the chemical potentials through
\begin{eqnarray}
    \mu_A(\mathbf{F}) - \mu_B(\mathbf{F}) = E^{A\rightarrow B}(N;\mathbf{F}) - {E}_{0}(N;\mathbf{F}) \,\,,
\end{eqnarray}
for all species pairs $(A,B)$, together with the constraint
\begin{equation}
\sum_A N_A\,\mu_A(\mathbf{F}) =E_0(N;\mathbf{F}),
\end{equation}
where $N_A$ is the number of atoms of species $A$ in the supercell and $\sum_{A} N_A = N$, and $E^{A\rightarrow B}(N;\mathbf{F})$ is the energy of the supercell when an atom of species $A$ is substituted by an atom of species $B$ ($A\neq B$).

To investigate the influence of mechanical loading, two representative deformation modes are considered: uniaxial loading and simple shear. For uniaxial deformation,
\begin{equation}
\mathbf{F} =\mathbf{I}+(\lambda-1)\,(\mathbf{e}_i\otimes\mathbf{e}_i),
\end{equation}
where $\lambda$ is the stretch ratio, corresponding to tensile loading for $\lambda>1$ and compressive loading for $\lambda<1$. Simple shear is described by
\begin{equation}
\mathbf{F} =\mathbf{I} + \gamma\, (\mathbf{e}_i\otimes\mathbf{e}_j),
\end{equation}
where $\gamma$ is the engineering shear strain. The associated volumetric strain,
\begin{equation}
\varepsilon_{\mathrm{vol}}=\det(\mathbf{F})-1,
\end{equation}
is equal to $\lambda-1$ for uniaxial loading and vanishes under simple shear.

\begin{figure}[ht!]\centering
\subfigure[uniaxial]{\includegraphics[keepaspectratio=true,width=0.4\textwidth]{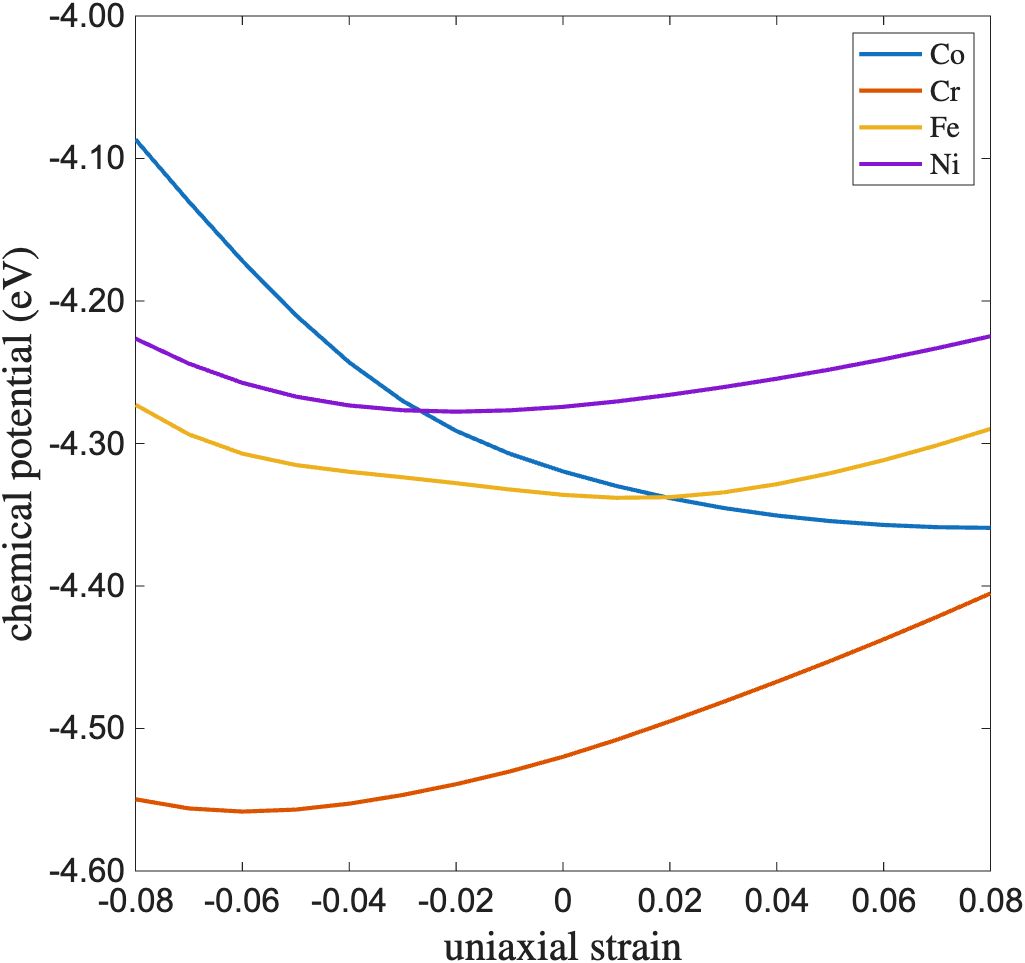}}
\subfigure[shear]{\includegraphics[keepaspectratio=true,width=0.4\textwidth]{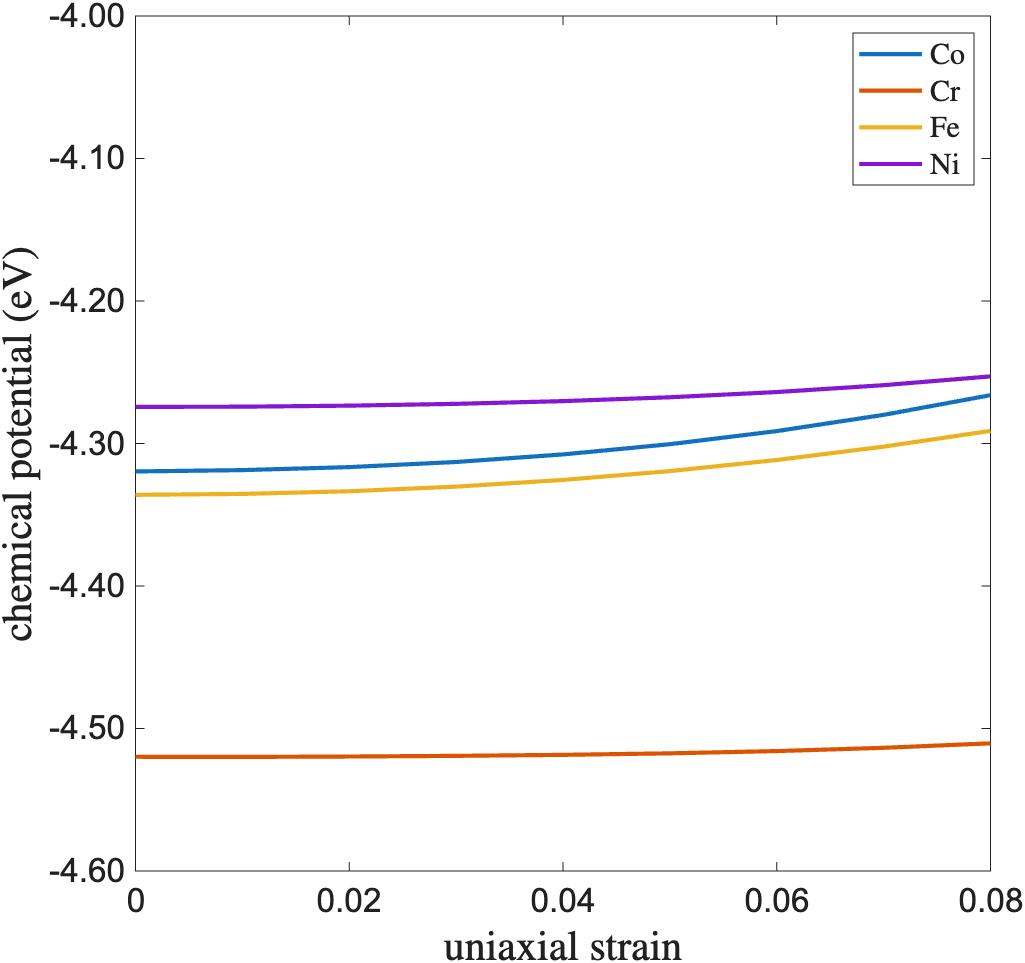}}
{\caption{Variation of average chemical potential with uniaxial and shear strain.}\label{Fig:ChemPot}}
\end{figure}

In this work, we use the equiatomic FCC CoCrFeNi high-entropy alloy for demonstration. The energies of the supercells are calculated using atomistic simulations, details of which are presented in Methods (Section \ref{sec:methods}). Figure~\ref{Fig:ChemPot} shows the variation of the average elemental chemical potentials with applied deformation.  Uniaxial loading produces substantial, species dependent changes in chemical potentials, reflecting the different responses of the constituent elements to volumetric deformation. In contrast, simple shear has a negligible effect on the chemical potentials, consistent with its nearly volume-preserving character.

\begin{figure}[ht!]\centering
\subfigure[CoCrFeNi]{\includegraphics[keepaspectratio=true,width=0.4\textwidth]{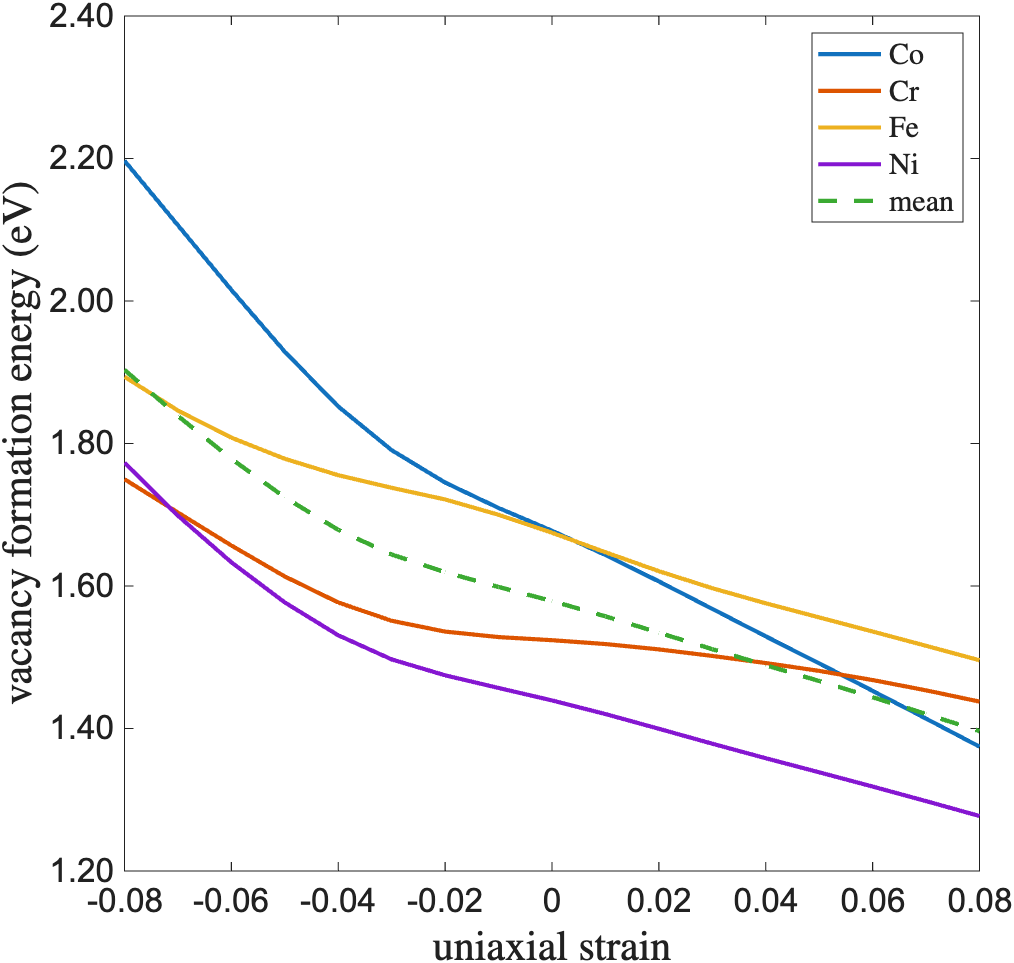}}
\subfigure[CoCrFeNi]{\includegraphics[keepaspectratio=true,width=0.4\textwidth]{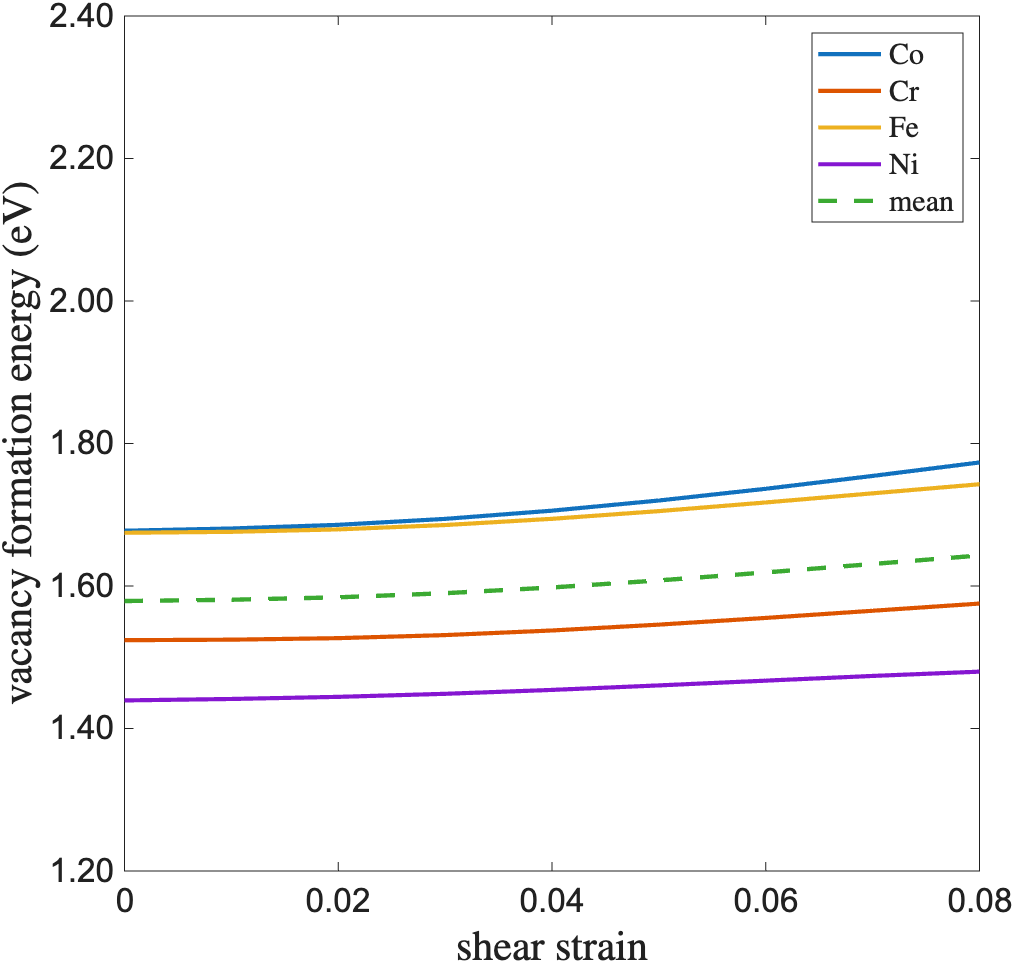}}
{\caption{Variation of vacancy formation energy with uniaxial and shear strain.}\label{Fig:Evf}}
\end{figure}

Figure~\ref{Fig:Evf} illustrates the effect of mechanical deformation on the average vacancy formation energy of each constituent species in the CoCrFeNi high entropy alloy. The vacancy formation energy exhibits a pronounced species dependence and decreases monotonically under tensile loading while increasing under compressive loading. In contrast, simple shear produces only negligible changes owing to its volume-preserving nature. Although these trends are qualitatively consistent with those reported for conventional FCC metals \cite{gavini2007vacancy,ghosh:2019}, the magnitude of the response is strongly species dependent, reflecting the chemically heterogeneous environment of high-entropy alloys. The dependence of the average vacancy formation energy on volumetric strain is well described by the empirical relation
\begin{equation}\label{Eq:fit}
E^{\mathrm{vf}}_A(\mathbf{F}) = a\left(1+\varepsilon_{\mathrm{vol}}\right)^{-n}+c,
\end{equation}
where $a$, $n$, and $c$ are species-dependent fitting parameters. The corresponding fits are shown in Table \ref{Table:fit}.

\begin{table}[h]
\centering
\begin{tabular}{lllll}
\hline\hline
vacancy type \,\,\,\, & a  & c & n \\
\hline
\hline
& &CoCrFeNi & \\
\hline
Co & $0.5896$ & $1.0749$ & $7.55$ \\
Cr & $0.0720$ & $1.4437$ & $17.29$ \\
Fe & $1.3427$ & $0.3237$ & $1.75$ \\
Ni & $0.2187$ & $1.2068$ &  $11.05$\\
mean & $0.3180$ & $1.2500$ & $8.36$ \\
 \hline
\end{tabular}\label{Table:fit}
\caption{Constants of the fit (\ref{Eq:fit}) for each species type as well as the mean.}
\end{table}

The observed scaling with volumetric strain indicates that the dominant effect of mechanical deformation on vacancy formation energetics arises from the associated change in crystal volume, while deviatoric deformation contributes comparatively little. Consequently, the empirical power-law relation provides a simple and computationally inexpensive description of the average strain dependence for each chemical species, making it attractive for continuum-scale constitutive and kinetic models. However, such an averaged representation neglects the intrinsic heterogeneity of high-entropy alloys. Owing to the chemically disordered local atomic environments, vacancy formation energies exhibit substantial site-to-site variations even for atoms of the same species, and these variations evolve under applied deformation. As a result, a single species-averaged curve cannot capture the statistical distribution of vacancy energetics or the coupling between local chemistry and mechanical loading. These limitations become particularly important for predicting vacancy-mediated phenomena, including diffusion, irradiation damage, and shock-induced defect evolution, where atomistically resolved vacancy energetics rather than average trends govern the underlying physics. This motivates the development of machine learning models capable of directly predicting strain-dependent vacancy formation energies from the local atomic environment.

\subsection{Permutation invariant neural network for deformation and chemical environment dependent vacancy energetics}\label{sec:GNN}

In chemically complex alloys, vacancy formation energetics depend sensitively on the local chemical and geometric environment surrounding the defect site \cite{Sugita2020,Mizuno2019,Esfandiarpour2019,Wang2022,Xu2022,spitaler2018perspectives,manzoor2021factors,linton2025mechanistic,Zhao2020,Zhou2022Nb}. Both vacancy formation and migration energies therefore exhibit broad distributions arising from local compositional disorder and lattice distortions. In the present work, vacancy formation is modeled as a supervised regression problem defined on vacancy-centered atomic neighborhoods. The objective is to learn a mapping from the local chemical environment and the deformation state to the vacancy formation energy, while preserving invariance under rigid-body rotations and permutations of neighboring atoms. 

In our work, we employ Deep Sets \cite{zaheer2017deep}, a permutation-invariant neural network architecture for learning functions on unordered sets. Each element is independently embedded into a latent space and combined using a symmetric aggregation operator, typically summation, before being processed by a second neural network. This construction guarantees invariance to permutations of the input while providing a flexible representation of the local atomic environment, making it well suited for atomistic machine-learning models.

For each vacancy site $i$, a local neighborhood is constructed as
\begin{equation}
\mathcal{N}_{i}=\{j \mid \|\mathbf{r}_j-\mathbf{r}_{i}\|\le r_c\},
\end{equation}
where $\mathbf{r}_{i}$ denotes the vacancy position, $\mathbf{r}_j$ is the position of neighboring atom $j$, $\mathbf{r}_i \neq \mathbf{r}_j$ and $r_c$ is a cutoff radius chosen to include the first three coordination shells surrounding the vacancy site.

The vacancy-centered neighborhood defines the local configurational state used by the surrogate model. In chemically disordered alloys, vacancy formation is governed predominantly by the immediate atomic environment surrounding the defect site because local compositional fluctuations modify bond topology, local lattice distortion, and short range chemical interactions. Restricting the representation to a finite cutoff radius therefore provides a localized approximation to the defect energetics while maintaining computational scalability.

Each neighboring atom $j\in\mathcal{N}_i$ is represented by a local feature vector
\begin{equation}
\mathbf{s}_{ij}=[z_j,\; r_{ij},\; \mathbf{a}_{ij}],
\end{equation}
where $z_j$ denotes the chemical species of atom $j$,
\begin{equation}
r_{ij} = \|\mathbf{r}_j-\mathbf{r}_{i}\|,
\end{equation}
and $\mathbf{a}_{ij}$ is a rotationally invariant angular descriptor characterizing the local geometric environment surrounding neighbor $j$.

The angular descriptor supplements the radial information by incorporating local coordination geometry around the vacancy. While pair distances characterize shell structure, angular correlations encode local distortions of the crystalline environment induced by chemical disorder and deformation. The use of averaged angular moments provides a compact rotationally invariant representation of local bond topology without introducing an explicit orientation frame.

The angular descriptor is defined as
\begin{equation}
\mathbf{a}_{ij}=
\left[
\frac{1}{N_j}\sum_{k\in\mathcal{N}_i,\,k\neq j}\theta_{i,jk},
\;
\frac{1}{N_j}\sum_{k\in\mathcal{N}_i,\,k\neq j}\cos\theta_{i,jk},
\;
\frac{1}{N_j}\sum_{k\in\mathcal{N}_i,\,k\neq j}\cos^2\theta_{i,jk}
\right],
\end{equation}
where $N_j$ is the number of neighboring atoms contributing to the angular environment of atom $j$, and
\begin{equation}
\theta_{i,jk} = \arccos\left(\frac{(\mathbf{r}_j-\mathbf{r}_{i})\cdot(\mathbf{r}_k-\mathbf{r}_{i})}{r_{ij} r_{ik}}\right)
\end{equation}
is the bond angle subtended by neighbors $j$ and $k$ relative to the vacancy site $i$ and $0\leq\theta_{i,jk}\leq\pi$.

Because the representation is constructed entirely from distances and angular correlations, it is invariant to rigid-body rotations and translations. Furthermore, the subsequent pooling operation renders the representation invariant to permutations of neighbor ordering. Such invariant geometric representations are widely used in atomistic machine learning frameworks for local energy prediction \cite{schutt2017schnet}.

Imposing rotational and permutation invariance is essential because the vacancy formation energy is a scalar thermodynamic quantity and therefore cannot depend on the arbitrary ordering of neighboring atoms or the orientation of the simulation cell. Constructing the representation from invariant geometric quantities ensures that physically equivalent environments map to identical latent representations.

The chemical identity is embedded through a learnable species embedding,
\begin{equation}
\mathbf{e}_j=\Phi(z_j),
\end{equation}
and $\mathbf{e}_j\in\mathbb{R}^{d_e}$. This embedding enables the network to learn a continuous latent representation of chemical identity directly from the training data. Instead of prescribing fixed descriptors for each elemental species, the embedding permits the model to infer chemically meaningful similarities and distinctions through optimization of the vacancy-energy regression objective. Mathematically, the embedding layer is a trainable matrix whose rows correspond to latent vectors associated with individual chemical species. The latent dimension $d_e$ was selected through validation studies. We performed a convergence study and observed that dimensions smaller than $64$ were not sufficient to distinguish chemically diverse local environments, whereas larger dimensions provided only marginal improvements at increased computational cost and model complexity. In our work, we choose $d_e=64$. Although the number of alloying species is relatively limited, the higher-dimensional latent space enables the network to represent nonlinear couplings among local chemistry, atomic geometry, and deformation state.

Next, $\mathbf{e}_j$ is concatenated with the radial and angular descriptors to form the initial node representation,
\begin{equation}
    \mathbf{h}^{(0)}_{ij}=[\mathbf{e}_j,\; r_{ij},\; \mathbf{a}_{ij}].
\end{equation}

Each node feature is processed through a shared nonlinear encoder,
\begin{eqnarray}
    \mathbf{h}^{(1)}_{ij} =\sigma(W_1\mathbf{h}^{(0)}_{ij}+b_1),
\end{eqnarray}
followed by
\begin{eqnarray}
    \mathbf{m}_{ij}=\sigma(W_2\mathbf{h}^{(1)}_{ij}+b_2),
\end{eqnarray}
Where $\sigma$ is the non-linear activation function, and the latent vector $\mathbf{m}_{ij}$ is the learned contribution of neighbor $j$ to the local vacancy environment. The encoder weights are shared across all neighboring atoms so that chemically and geometrically equivalent local motifs are processed consistently, independent of neighbor ordering across different vacancy configurations. The encoder operates on a $68$-dimensional input feature vector constructed from the concatenation of the chemical-species embedding ($64$-dimensional), radial descriptor ($1$-dimensional), and angular descriptor ($3$-dimensional) associated with each neighboring atom. The first encoder layer projects this $68$-dimensional input into a $128$-dimensional latent feature space through a learned affine transformation followed by a nonlinear activation. The increased latent dimensionality provides sufficient representational capacity for the network to capture nonlinear interactions between local chemistry, coordination geometry, and deformation-induced distortions surrounding the vacancy site. A second nonlinear transformation is then applied within the same $128$-dimensional latent space to produce the encoded neighbor representation $\mathbf{m}_j$. Maintaining a fixed $128$-dimensional latent representation across successive encoder layers preserves a consistent feature hierarchy while providing sufficient representational capacity to capture the complex coupling between local chemistry, coordination geometry, and vacancy energetics.

We construct a permutation-invariant representation of the local environment by summation over the learned neighbor descriptors,
\begin{equation}
\bm{\xi}_i=\sum_{j\in\mathcal{N}_i}\mathbf{m}_{ij},
\end{equation}
thereby ensuring invariance to the ordering of atoms within the local environment. For computational efficiency, neighborhoods are embedded into fixed-size tensors of length $N_{\max}$ using zero-padding. A binary mask $M_j\in\{0,1\}$ is applied to exclude padded entries, yielding
\begin{equation}
\bm{\xi}_i=\sum_{j=1}^{N_{\max}} M_j \mathbf{m}_{ij}.
\end{equation}
The vector $\bm{\xi}_i$ can be interpreted as a learned representation of the local chemical environment surrounding the vacancy at site $i$. Because the summation operation is permutation invariant, the resulting descriptor depends only on the statistical and geometric characteristics of the neighborhood rather than the indexing of atoms within the local environment.

To incorporate the effect of deformations, we use the principal invariants of the right Cauchy--Green tensor, $\mathbf{C}=\mathbf{F}^{T}\mathbf{F}$, as the deformation descriptor. The invariants $I_1$, $I_2$, and $I_3$ are given by
\begin{eqnarray}
    I_1 &=& \mathrm{tr}(\mathbf{C})\,\,, \\
    I_2 &=& \frac{1}{2} \left[ (\mathrm{tr}\,\mathbf{C})^2 -\mathrm{tr}(\mathbf{C}^2) \right] \,\,, \\
    I_3 &=& \det(\mathbf{C}) \,\,.
\end{eqnarray}
These principal invariants of the right Cauchy--Green tensor ensure objectivity of the mechanical-strain descriptor \cite{ogden1997non}. Direct use of the deformation-gradient components would introduce an artificial dependence on the coordinate-system orientation, whereas the invariant representation isolates physically meaningful measures of volumetric and distortional deformation. This objectivity is particularly important when comparing vacancy energetics across multiple loading paths and heterogeneous local environments.

Next, the principal invariants are concatenated with the deformation descriptor to obtain the coupled environment-strain descriptor $\bm{\eta}_i \in\mathbb{R}^{131}$,
\begin{equation}\label{Eq:EnvStr}
    {\bm{\eta}_i} = [\bm{\xi}_i,\;I_1,I_2,I_3] \,\,.
\end{equation}
This is subsequently passed through a dense neural network layer 
\begin{equation}
    \mathbf{h}^{(2)}_i = \mathrm{\sigma}(W_3\bm{\eta}_i+b_3),
\end{equation}
followed by a linear output projection yielding the prediction of the normalized vacancy formation energy
\begin{equation}
   \hat{{E}}_{vf,{\mathrm{pred}}}^i = W_4\mathbf{h}_i^{(2)}+b_4.
\end{equation}
This sequence defines a nonlinear transformation from the $131$-dimensional latent feature space to a $128$-dimensional hidden representation, followed by projection onto a scalar output.

We remark that the combined latent representation (Equation \ref{Eq:EnvStr}) couples the local chemical environment with the local mechanical state, enabling the surrogate to learn nontrivial interactions between strain and chemical disorder. In chemically complex alloys, deformation alters local bond lengths, coordination asymmetry, and lattice distortion in a composition-dependent manner, leading to strongly heterogeneous vacancy formation behavior across nominally equivalent lattice sites.

The training target is the normalized vacancy formation energy,
\begin{eqnarray}
     \hat{ {E}}_{vf}^i = \frac{{E}_{vf}^i-\mu_{\mathrm{train}}}{\sigma_{\mathrm{train}}},
\end{eqnarray}
where $\mu_{\mathrm{train}}$ and $\sigma_{\mathrm{train}}$ are the mean and standard deviations of the training data. The model is optimized using the smooth $L_1$ Huber loss
\begin{eqnarray}
    \mathcal{L}(\delta)=
\begin{cases}
\dfrac{1}{2}\delta^2,
&
|\delta|<1,
\\[6pt]
|\delta|-\dfrac{1}{2},
&
|\delta|\ge 1,
\end{cases}
\end{eqnarray}
where
\begin{eqnarray}
    \delta =  \hat{{E}}_{vf,{\mathrm{pred}}}^i- \hat{ {E}}_{vf}^i \,\,.
\end{eqnarray}
The Huber loss provides a regression objective that is less sensitive to large residuals than mean-squared-error optimization while retaining smooth gradient behavior in the vicinity of the optimum \cite{huber1964robust}. This is particularly advantageous for vacancy energetics in chemically complex alloys, where statistically rare local environments may generate unusually high or low formation energies relative to the typical range of the distribution. Because the identity of the removed atomic species is not imposed explicitly as an independent input variable, the network instead learns the effective energetic cost of vacancy formation directly from the surrounding atomic environment encoded within the permutation invariant representation.

Finally, the proposed framework differs from previous machine-learning models for defect and vacancy energetics that employ descriptor-based representations, graph neural networks, or composition–property learning strategies \cite{manzoor2021ml, wang2022gbvfe, Zhou2022H, linton2025mechanistic, tan2025lae} in several important respects. First, the framework adopts a vacancy-centered local representation based on permutation- and rotation-invariant geometric descriptors, providing an objective description of the local atomic environment. Second, finite mechanical deformation is incorporated explicitly through objective invariants of the local deformation gradient, enabling the framework to capture coupling between local strain fields and vacancy formation energetics. Third, the removed atomic species is not introduced as an independent categorical feature. Instead, its influence is inferred implicitly from the surrounding local chemical environment encoded in the vacancy-centered representation.

\subsection{Rotational invariance of descriptors}\label{Sec:RotInvar}
In this section, we prove the rotational invariance of the descriptors described in the previous section. Let the vacancy be at site $i$ with position vector $\mathbf{r}_{i}$, and vacancy-centered relative position vectors of the atom at site $j$ be defined as
\begin{equation}
\mathbf{u}_j=\mathbf{r}_j-\mathbf{r}_{i}.
\end{equation}
Then the radial descriptor is
\begin{equation}
r_{ij}=\|\mathbf{u}_{ij}\|,
\end{equation}
and the bond angle subtended by neighbors $j$ and $k$ relative to the vacancy is
\begin{equation}
\theta_{i,jk}=\arccos\left(\frac{\mathbf{u}_{ij}\cdot\mathbf{u}_{ik}}{\|\mathbf{u}_{ij}\|\,\|\mathbf{u}_{ik}\|}\right).
\end{equation}
To show rotational invariance, consider an arbitrary rigid-body rotation represented by an orthogonal matrix
\begin{equation}
\mathbf{Q}\in SO(3), \qquad \mathbf{Q}^T\mathbf{Q}=\mathbf{I}, \qquad \det \mathbf{Q}=1.
\end{equation}
Under rotation, $\mathbf{u}_{ij} \rightarrow \mathbf{u}_{ij}'=\mathbf{Q}\mathbf{u}_{ij} $.

The transformed radial distance becomes $r_{ij}'=\|\mathbf{u}_{ij}'\|=\|\mathbf{Q}\mathbf{u}_{ij}\|$. 

Again using orthogonality of $\mathbf{Q}$, it is easy to see that 
\begin{equation}
\|\mathbf{Q}\mathbf{u}_{ij}\|^2=
(\mathbf{Q}\mathbf{u}_{ij})^T(\mathbf{Q}\mathbf{u}_{ij}) =
\mathbf{u}_{ij}^T\mathbf{Q}^T\mathbf{Q}\mathbf{u}_{ij} =
\mathbf{u}_{ij}^T\mathbf{u}_{ij} =
\|\mathbf{u}_{ij}\|^2 \,\,.
\end{equation}
Therefore, $r_{ij}'=r_{ij}$. Hence, the radial descriptor is rotationally invariant.

Next, we show that the angular descriptors are rotationally invariant. The transformed angle satisfies
\begin{equation}
\cos\theta_{i,jk}'=\frac{\mathbf{u}_{ij}'\cdot\mathbf{u}_{ik}'}{\|\mathbf{u}_{ij}'\|\,\|\mathbf{u}_{ik}'\|}.
\end{equation}
Substituting $\mathbf{u}_{ij}'=\mathbf{Q}\mathbf{u}_{ij}$, we obtain
\begin{equation}
\cos\theta_{i,jk}'= \frac{(\mathbf{Q}\mathbf{u}_{ij})\cdot(\mathbf{Q}\mathbf{u}_{ik})}{\|\mathbf{Q}\mathbf{u}_{ij}\|\,\|\mathbf{Q}\mathbf{u}_{ik}\|}.
\end{equation}
Using $(\mathbf{Q}\mathbf{u}_{ij})\cdot(\mathbf{Q}\mathbf{u}_{ik}) =
\mathbf{u}_{ij}^T\mathbf{Q}^T\mathbf{Q}\mathbf{u}_{ik} =\mathbf{u}_{ij}\cdot\mathbf{u}_{ik}$,   $\|\mathbf{Q}\mathbf{u}_{ij}\|=\|\mathbf{u}_{ij}\|$, and $\|\mathbf{Q}\mathbf{u}_{ik}\|=\|\mathbf{u}_{ik}\|$, we obtain
\begin{equation}
\cos\theta_{i,jk}'
=
\frac{\mathbf{u}_{ij}\cdot\mathbf{u}_{ik}}
{\|\mathbf{u}_{ij}\|\,\|\mathbf{u}_{ik}\|}
=
\cos\theta_{jk}.
\end{equation}
Since $\arccos(\cdot)$ is a scalar-valued function, it follows that $\theta_{i,jk}'=\theta_{i,jk}$. Consequently, every component of $\mathbf{a}_{ij}$ remains invariant under rotation. It therefore follows that $\mathbf{h}_{ij}^{(0)}=[\mathbf{e}_j,r_{ij},\mathbf{a}_{ij}]$ is likewise rotationally invariant. Because the encoder applies identical nonlinear transformations independently to each node, the resulting learned representation $\bm{\xi}_i$ also preserves rotational invariance. Additionally, the principal invariants of $\mathbf{C}$ are rotationally invariant \cite{ogden1997non,anand2020continuum}.

Since all descriptors remain unchanged under rigid-body rotations, the predicted vacancy formation energy is also rotationally invariant.

\subsection{Neural network prediction}\label{Sec:NNPred}
In this section, we first assess the accuracy of the neural network in predicting the site-specific vacancy formation energy, and then use the network to calculate the effect of stress on site-specific vacancy formation enthalpy, formation entropy, and site-specific vacancy concentration.

\begin{figure}[ht!]\centering
\subfigure[parity]{\includegraphics[keepaspectratio=true,width=0.45\textwidth]{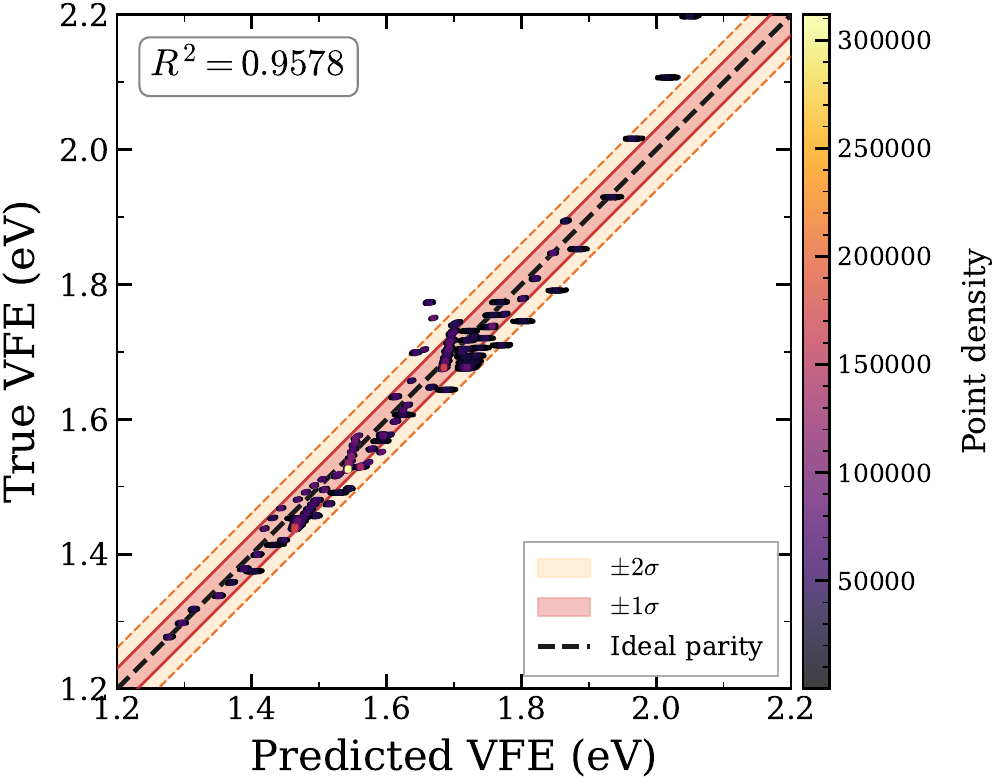}\label{Fig:parity}}
\subfigure[vacancy formation energy]{\includegraphics[keepaspectratio=true,width=0.5\textwidth]{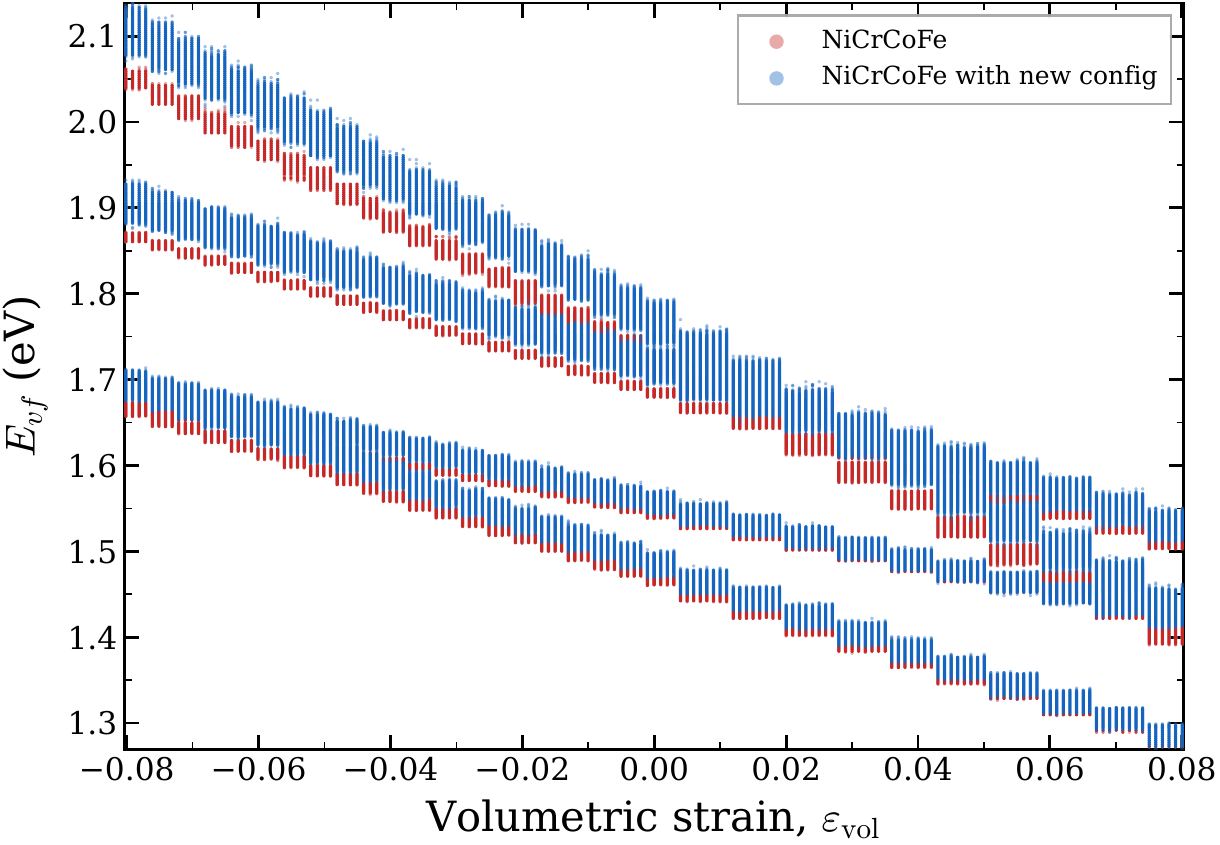}\label{Fig:VFEPred} }
{\caption{(a) shows the true versus predicted values of vacancy formation energy. (b) shows the predicted values of vacancy formation energy versus volumetric strain. The ``new configuration" refers to a randomly generated atomic configuration not used for training.}}
\end{figure}

Figure \ref{Fig:parity} compares the predicted vacancy formation energies with the corresponding reference values over the full energy range considered. The majority of predictions lie close to the ideal parity line, demonstrating that the model reproduces the principal variations in vacancy formation energy without appreciable systematic bias. The high coefficient of determination ($R^2=0.9578$) further confirms the predictive performance of the model and suggests that the learned representations capture the local chemical and geometric features controlling vacancy energetics. The largest deviations occur near the extremes of the energy distribution, where the available data are comparatively sparse, and the configurational environments are more complex. Taken together, these results show that the permutation invariant neural network surrogate provides an accurate and computationally efficient approach for large-scale prediction of strain-dependent vacancy formation energies in chemically complex alloys.

In Figure \ref{Fig:VFEPred}, we plot the site-speccific vacancy formation energy versus volumetric strain for different chemical environments as predicted by the trained neural network. In order to assess the ability of the model to generalize beyond the specific atomic realization encountered during training, we consider two distinct supercell configurations, namely the original CoCrFeNi structure used during training and a second independently generated configuration. From Figure \ref{Fig:VFEPred}, a clear monotonic dependence on volumetric strain is observed, decreasing under tensile dilation and increasing under compression. This behavior is consistent with the trends discussed in Section~\ref{Sec:deformationVFE} and arises from the coupling between vacancy formation energetics and the local volumetric work associated with defect creation. The responses corresponding to the original CoCrFeNi configuration and the independently generated configuration remain nearly coincident across the full strain range, indicating that the strain dependence is largely insensitive to variations in the underlying atomic realization. The presence of multiple closely spaced curves indicates that distinct local chemical environments yield different vacancy formation energies, while exhibiting a similar dependence on applied strain, as reflected in nearly identical slopes. These results demonstrate that volumetric deformation primarily shifts the absolute vacancy formation energy while preserving the energetic hierarchy imposed by chemical disorder.

In mechanically loaded solids, defect formation is governed not only by changes in internal energy associated with electronic and configurational rearrangements, but also by the mechanical work performed against the applied stress field. Under finite hydrostatic pressure, the relevant thermodynamic potential is therefore the enthalpy, which accounts explicitly for the energetic contribution associated with volume changes.

The formation of a vacancy produces a local volumetric distortion that couples to the applied pressure through a $P\Delta V$ contribution. This term represents the mechanical work exchanged with the surroundings during defect formation and is not included in the internal energy alone. Accordingly, the vacancy formation enthalpy is evaluated using the hydrostatic pressure $P=-\frac{1}{3}\tr{\bm{\sigma}}$. Under the sign convention adopted here, negative pressure denotes hydrostatic tension, whereas positive pressure denotes hydrostatic compression. The site-specific vacancy formation enthalpy $ H_{vf}^i$ as a function of pressure is obtained from the site-specific vacancy formation energy $ E_{vf}^i$ using a Legendre transform
\begin{eqnarray}
    H_{vf}^i(P,\bm{\xi}_i) = E_{vf}^i(\mathbf{F},\bm{\xi}_i) + P\, V_{vf}^i \,\,
\end{eqnarray}
where $ V_{vf}^i$ is the vacancy formation volume and is obtained as the difference between $V_{defect}^i$, the volume of the cell with a defect at site $i$, and the scaled volume of the perfect cell $V_{perfect}$ without defects. This is given by
\begin{eqnarray}
     V_{vf}^i = V_{defect}^i-\frac{N-1}{N}V_{perfect}  \,\,.
\end{eqnarray}

\begin{figure}[ht!]\centering
{\includegraphics[keepaspectratio=true,width=0.5\textwidth]{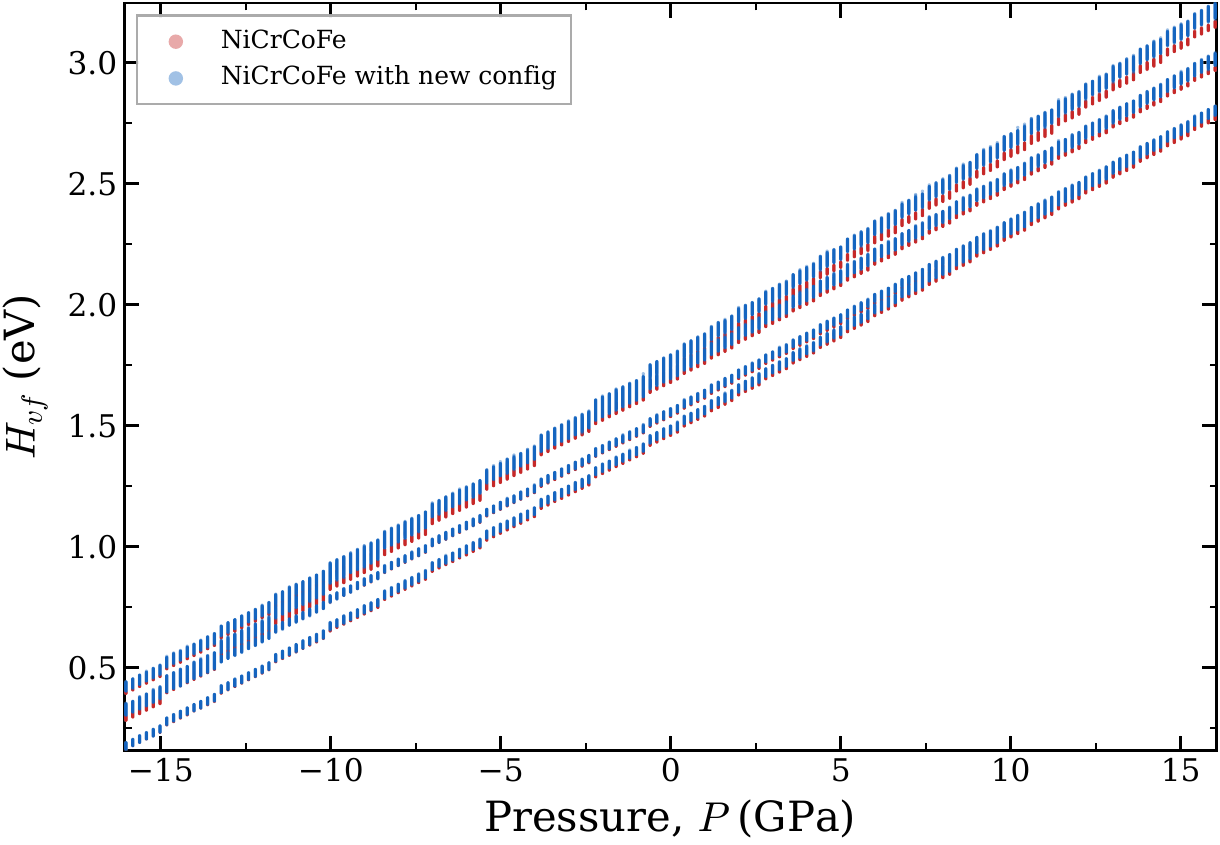}}
{\caption{The effect of pressure on vacancy formation enthalpy. The ``new configuration" refers to a randomly generated atomic configuration not used for training.}\label{Fig:enthalpy}}
\end{figure}

Figure \ref{Fig:enthalpy} shows the vacancy formation enthalpy as a function of pressure for different local environments. From this figure, we see that the vacancy formation enthalpy increases with increasing applied pressure; consequently, tensile hydrostatic stress lowers the formation enthalpy, whereas compressive hydrostatic stress increases it. The close overlap between the atomic configurations used for training and the new configuration demonstrates that the pressure dependence is largely insensitive to the specific atomic realization. Moreover, chemically distinct vacancy environments preserve their relative energetic ordering under both tensile and compressive loading. These results indicate that hydrostatic pressure acts primarily as a systematic shift in the vacancy formation enthalpy, while leaving the hierarchy among local chemical environments largely unchanged.

The equilibrium site-specific vacancy occupancy probability $c_v^i$ also depends on temperature, pressure, and chemical environments. This is given by
\begin{eqnarray}
    c_v^i(P,\bm{\xi}_i;T) = \exp\left({\frac{S_{vf}^i(P,\bm{\xi}_i)}{k_B}}\right)\exp\left(\frac{-H_{vf}^i(P,\bm{\xi}_i)}{k_BT}\right) \,\,,
\end{eqnarray}
Where $T$ is the absolute temperature, $k_B$ is the Boltzmann constant, and $S_{vf}^i$ denotes the site-specific vacancy formation entropy, defined as the entropy change associated with the creation of a vacancy in a crystalline lattice. \textcolor{black}{The vacancy formation free energy may, in general, contain vibrational, electronic and, where relevant, magnetic entropy contributions. In the present room-temperature analysis, these contributions are not evaluated explicitly. Instead, we focus on the configurational entropy associated with the distribution of vacancy occupation probabilities among chemically nonequivalent lattice sites. We assume that site to site variations in the omitted entropy contributions are small relative to the corresponding variations in vacancy formation enthalpy and therefore do not qualitatively alter the pressure and chemistry dependent trends considered here. Accordingly, the reported results should be interpreted as enthalpy-based estimates of vacancy propensity and configurational site occupation statistics, rather than as quantitatively complete vacancy formation free energies or absolute equilibrium vacancy concentrations.}

The configurational contribution to the site-specific vacancy formation entropy is calculated from the conditional vacancy-location probability distribution over distinct lattice sites. At thermal equilibrium, the conditional probability of a vacancy at location at site $i$ is governed by Boltzmann statistics,
\begin{equation}
p_i(P,{\bm{\xi}_i}) = \frac{\exp\!\left(-H_{vf}^i/{k_B T}\right)}{\displaystyle \sum_{i=1}^N \exp\!\left(-H_{vf}^i/k_B T\right)} \,\,,
\label{eq:pi}
\end{equation}
where the dependence of the probability $p_i$ on pressure and the chemical environment arises from the corresponding dependence of the vacancy formation enthalpy at site $i$ on these variables. The denominator in Equation \ref{eq:pi} is the configurational partition function associated with the ensemble of vacancy formation sites. The configurational contribution to the vacancy formation entropy is then obtained from the Shannon entropy associated with the probability distribution $\{p_i\}$,
\begin{equation}
S_{vf,\mathrm{conf}}^i(P)  = -k_B \sum_{i=1}^N p_i \ln p_i \,\,,
\label{eq:sconf_discrete}
\end{equation}
where the local chemical environment descriptor $\bm{\xi}_i$ does not appear explicitly on the left-hand side because its contribution is implicitly incorporated through the summation over all atomic sites. Equation~\eqref{eq:sconf_discrete} quantifies the degree of statistical heterogeneity in vacancy occupation arising from chemical disorder. 

\begin{figure}[ht!]\centering
\subfigure[probability of vacancy occupation]{\includegraphics[keepaspectratio=true,width=0.5\textwidth]{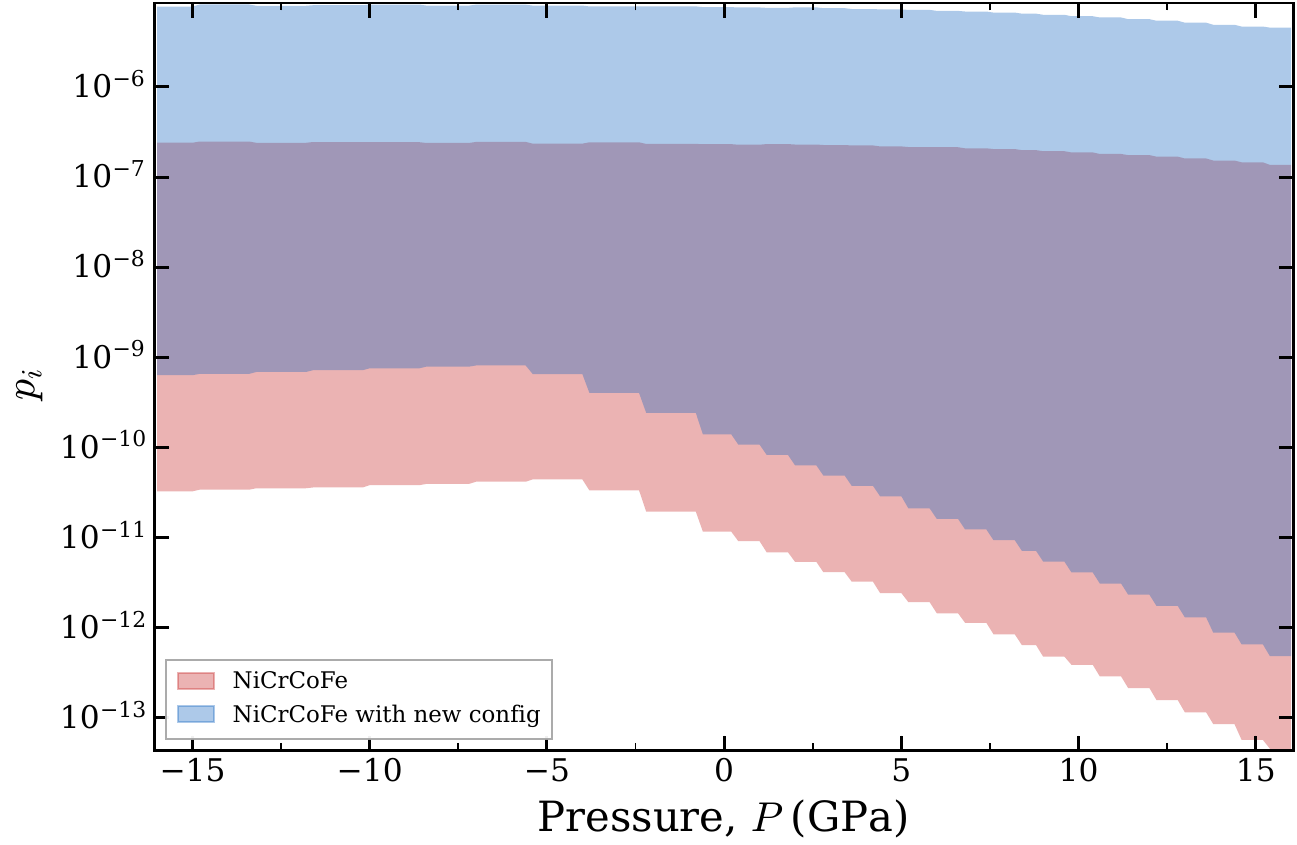}\label{Fig:pi}}
\subfigure[entropy]{\includegraphics[keepaspectratio=true,width=0.5\textwidth]{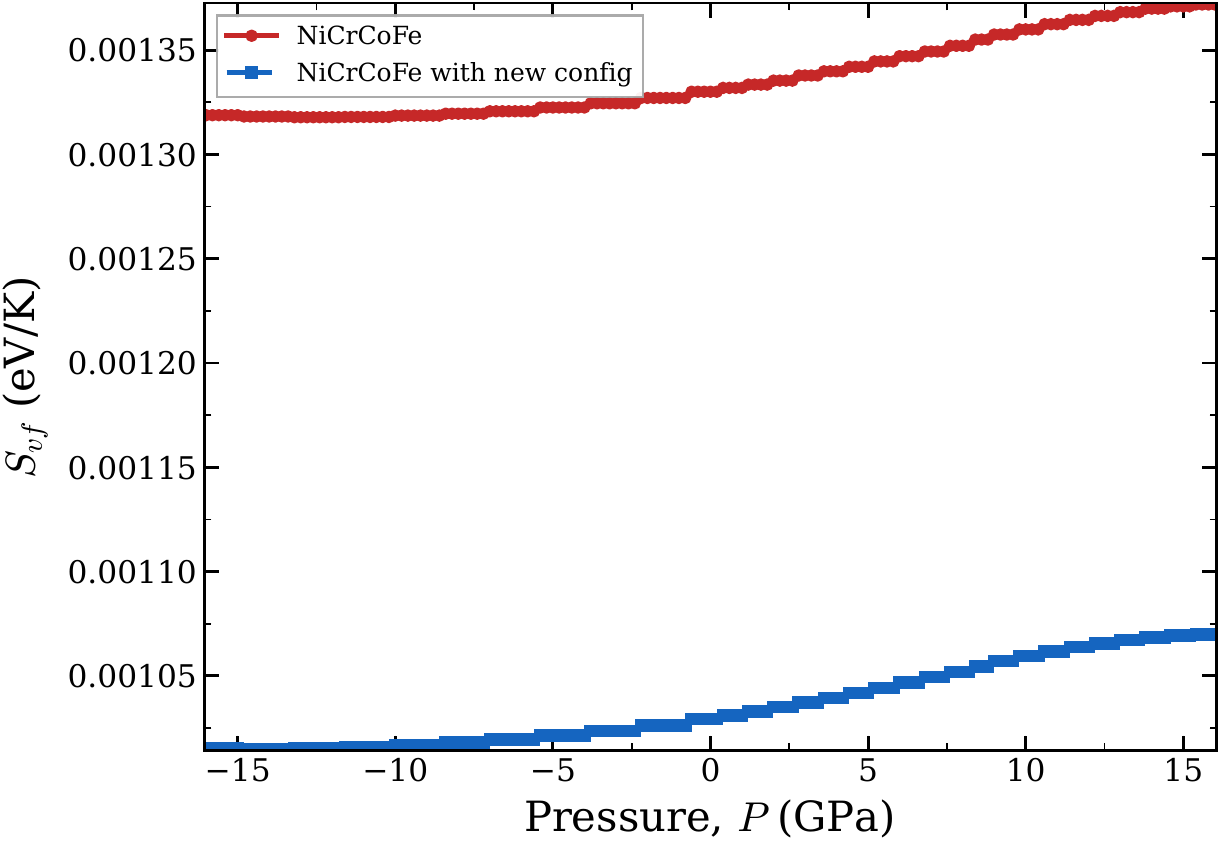}\label{Fig:SvfNN}}
{\caption{(a) shows the true versus predicted values of vacancy formation energy. (b) shows the predicted values of vacancy formation energy versus volumetric strain}}
\end{figure}

Figure \ref{Fig:pi} shows the influence of hydrostatic pressure on $p_i$, for the CoCrFeNi alloy with $P$ ranging from = -$16\text{ GPa}$ to $16\text{GPa}$. From this figure, we observe that $p_i$ exhibits a monotonic, exponentially decreasing trend with increasing hydrostatic pressure for both investigated configurations. Under severe tensile loading ($-16\text{ GPa}$), the dilation of the lattice lowers the thermodynamic barrier for atom removal, resulting in a maximum probability of $\approx 10^{-6}$. Conversely, compressive hydrostatic states up to $16\text{ GPa}$ significantly penalizes the local volumetric strain required to accommodate a vacancy, decreasing $p_i$ to $\approx 10^{-13}$. The overall pressure dependence of $p_i$ is consistent across different atomic configurations, although its value varies between configurations.

Figure \ref{Fig:SvfNN} shows the variation of the configurational contribution to the vacancy formation entropy with hydrostatic pressure. The configurational entropy increases monotonically under compression and decreases under tension. Although the two atomic configurations yield different absolute values, they exhibit the same overall pressure-dependent trend.

\begin{figure}[ht!]\centering
{\includegraphics[keepaspectratio=true,width=0.5\textwidth]{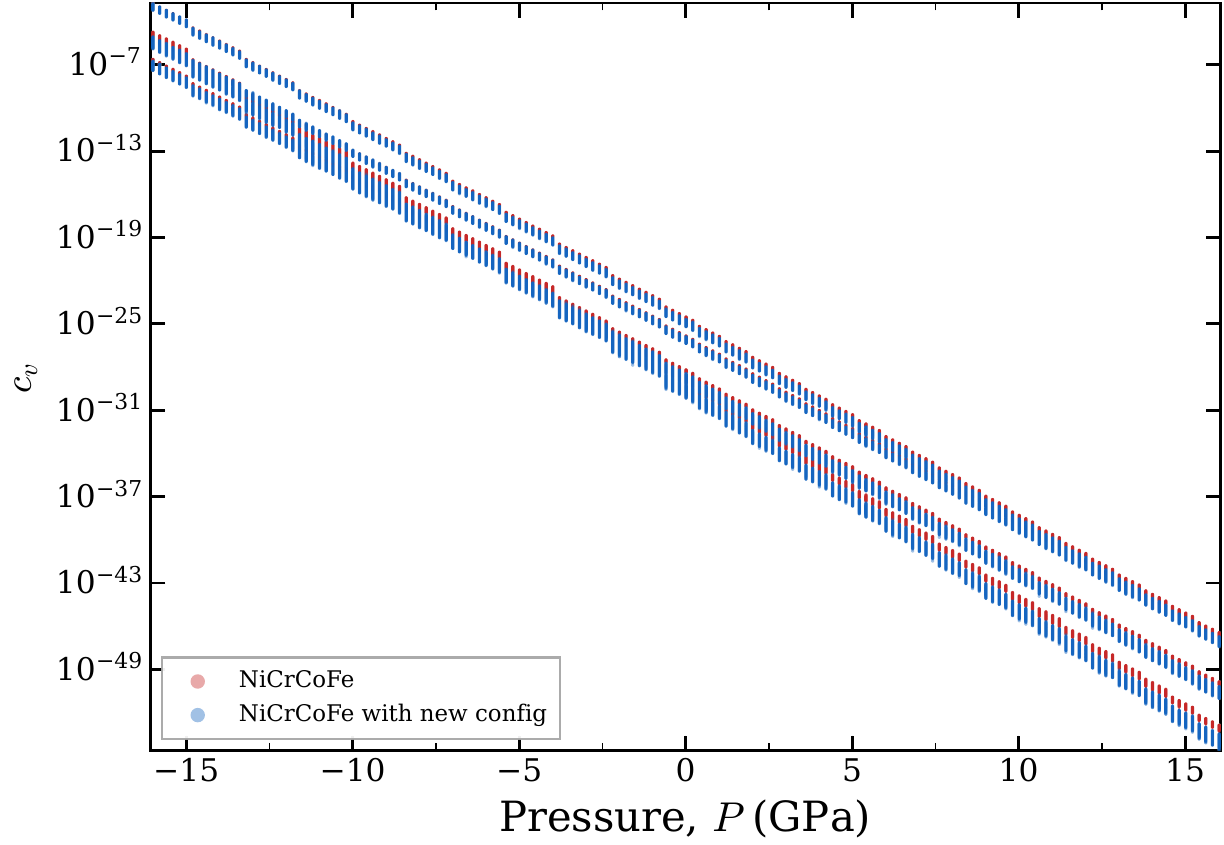}}
{\caption{The effect of pressure on the concentration of vacancy at temperature $T=300$K.}\label{Fig:conc}}
\end{figure}

Figure \ref{Fig:conc} shows the dependence of the equilibrium site-specific vacancy concentration on hydrostatic pressure. A pronounced thermodynamic coupling between volumetric stress and vacancy populations is evident. The equilibrium site-specific vacancy concentration decreases monotonically with increasing hydrostatic compression and increases under hydrostatic tension. This behavior arises from the corresponding pressure dependence of the vacancy formation enthalpy: compression increases the energetic cost of vacancy formation, thereby suppressing vacancy populations, whereas tension lowers the formation enthalpy and promotes vacancy formation, consistent with the trends shown in Figure \ref{Fig:VFEPred}.

The nearly parallel pressure-dependent responses obtained for the two atomic configurations indicate that the sensitivity of vacancy concentration to hydrostatic loading is largely independent of the specific atomic realization. However, the offset between the two curves demonstrates that distinct local chemical environments produce different equilibrium vacancy concentrations while exhibiting a similar pressure response. Collectively, these results show that hydrostatic pressure provides an effective thermodynamic means of tuning vacancy populations in chemically complex alloys without substantially altering the relative influence of local chemical disorder.
 
\section{Remarks and outlook}\label{sec:remarks}
In this work, we developed an atomistically informed machine-learning framework for predicting strain-dependent vacancy formation energies in FCC high-entropy alloys. By combining large-scale atomistic simulations with a vacancy-centered local representation, the proposed model captures the coupled effects of local chemical disorder and mechanical deformation on vacancy energetics. The results demonstrate that volumetric deformation is the primary driver of the average variation in vacancy formation energy, whereas shear deformation has only a minor influence. At the same time, substantial site-to-site variability persists even under identical macroscopic loading, highlighting the intrinsically statistical nature of vacancy energetics in chemically disordered alloys. These findings show that species-averaged constitutive relations are insufficient to describe vacancy thermodynamics in high-entropy alloys and motivate the need for atomistically resolved predictive models.

The proposed framework provides an efficient alternative to large-scale atomistic simulations while retaining the ability to resolve local chemical and mechanical effects. Because the model predicts vacancy formation energies directly from local atomic environments, it is well suited for high-throughput screening of defect energetics across broad composition and deformation spaces. Such capabilities provide a foundation for incorporating stress-dependent vacancy thermodynamics into multiscale descriptions of diffusion, irradiation damage, phase transformations, and dynamic failure in chemically complex alloys.

Several opportunities exist for extending the present framework. The current study is restricted to monovacancy formation in an FCC CoCrFeNi alloy under homogeneous deformation. Future work will generalize the methodology to multicomponent alloy systems with varying compositions, other crystal structures, including BCC and HCP high-entropy alloys, and more complex loading paths involving multiaxial stress states and finite deformations. An important direction is the extension of the framework to predict migration barriers, vacancy clusters, interstitials, dislocation--vacancy interactions, and other defect energetics within a unified machine-learning framework. Coupling these atomistically informed predictions with kinetic Monte Carlo, phase-field, crystal plasticity, and concurrent atomistic--continuum methods would further enable quantitative simulations of defect evolution across experimentally relevant length and time scales. More broadly, the proposed framework illustrates how physics-informed local representations and machine learning can accelerate the prediction of defect energetics in chemically complex materials, providing a pathway toward scalable defect modeling and the inverse design of structural alloys for extreme environments.

\section{Methods}\label{sec:methods}

\subsection{Atomistic simulations}

Atomistic simulation cells representing equiatomic FCC CoCrFeNi random solid solutions were generated using the OPERA package within the random solid-solution approximation \cite{anand2023order}. Classical molecular statics calculations were performed using the Large-scale Atomic/Molecular Massively Parallel Simulator (LAMMPS) \cite{thompson2022lammps} with the embedded-atom method (EAM) potential developed by Farkas and Caro \cite{farkas2018model}. Vacancy-containing configurations were constructed by removing a single atom from the relaxed random solid solution, followed by relaxation of the atomic coordinates and simulation cell. Finite deformation was subsequently imposed by applying the prescribed deformation gradient to the relaxed atomic configuration \cite{ghosh2024violation}. The vacancy formation energy under deformation was then computed by relaxing the atomic positions while keeping the deformed simulation cell fixed, thereby corresponding to a prescribed macroscopic deformation.

The calculated equilibrium lattice constant is $3.54$ Angstroms, in excellent agreement with the experimentally measured value of $3.57$ Angstroms \cite{huang2019element}, previous atomistic simulations ($3.54$–$3.55$ Angstroms) \cite{Nitol,choi2018understanding}, and first-principles calculations ($3.56$ Angstroms) \cite{ge2018effect}. For the stress-free configuration, the predicted vacancy formation energies of CoCrFeNi range from $1.43$ to $1.67$ eV, which fall within the range of values reported in the literature. Previous studies have reported vacancy formation energies between $1.55$ and $2.25$ eV, with a broader range of $0.72$–$3.09$ eV reflecting variations arising from different computational methods and local chemical environments \cite{CHEN2018355}.

\subsection{Details of training and computational resources}
\label{sec:training}
The dataset comprised $500,000$ vacancy formation energies sampled across diverse local chemical environments in the CoCrFeNi high-entropy alloy for each point in the discretized strain space. The strain space was discretized into $17$ uniaxial-strain states and $17$ simple shear-strain states, thereby enabling systematic sampling of the coupled effects of chemical disorder and mechanical deformation on vacancy energetics. We define the local chemical environment as the first three nearest-neighbor coordination shells surrounding the vacancy. 

All atomistic simulations and neural-network training were performed on the Oak Ridge Leadership Computing Facility Frontier supercomputer \cite{frontier2022} using GPU-accelerated compute nodes consisting of one 64-core AMD EPYC processor coupled to four AMD Instinct MI250X accelerators. Molecular dynamics simulations employed domain-decomposed parallel execution with GPU-accelerated force evaluation and neighbor-list construction, enabling efficient generation of the large ensemble of deformation configurations used for model training. The neural network was implemented in PyTorch and trained using distributed multi-GPU optimization with batched data resident on accelerator memory during backpropagation.

The neural regressor was trained using the AdamW optimizer \cite{loshchilov2017decoupled} and rectified linear unit (ReLU) \cite{nair2010rectified} as the nonlinear activation function. An initial learning rate of $10^{-3}$ and a weight decay coefficient of $10^{-4}$ were used. Model optimization was performed for 1000 epochs using a batch size of 256 and the smooth $L_1$ (Huber) loss function.  The dataset was partitioned into training, validation, and test subsets using a fixed random seed and a $70\%$, $15\%$, $15\%$ split. To improve scalability for large datasets, the training data were stored in cached array files and accessed through memory-mapped loading. Distributed data parallelism was used to accelerate training across multiple compute devices. Model checkpoints were saved during training corresponding to the best validation mean absolute error (MAE), best validation root mean square error (RMSE), and lowest validation loss.

\begin{figure}[ht!]
    \centering
    {\includegraphics[keepaspectratio=true,width=0.5\textwidth]{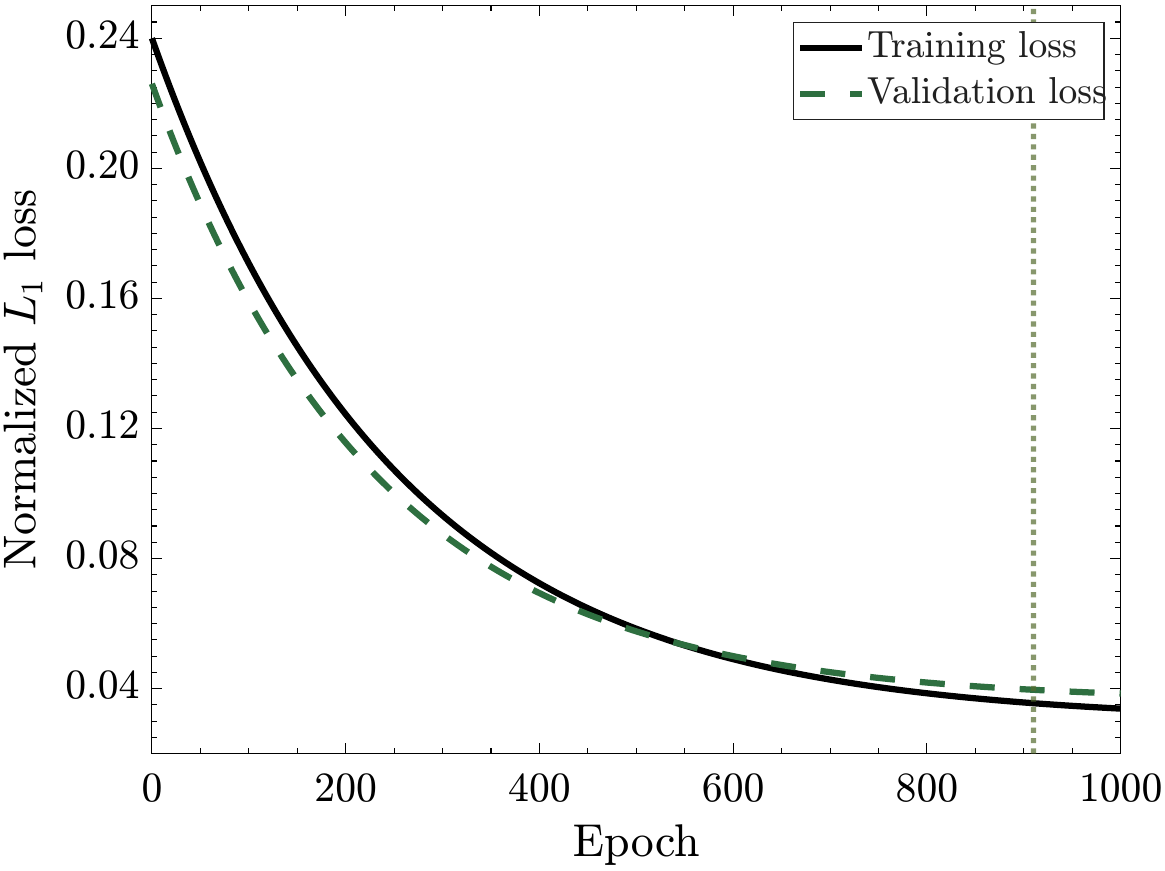}}
    \caption{
    Training history of the invariant local-environment neural regressor used for vacancy formation energy prediction. The normalized Smooth $L_1$ training and validation loss are shown as a function of epoch. The lowest validation loss achieved during training is $0.034$ at epoch $908$.
    }
    \label{fig:loss_curve_only}
\end{figure}

Figure~\ref{fig:loss_curve_only} presents the evolution of the training and validation loss during optimization. The normalized loss decreases monotonically for both datasets throughout training, indicating stable convergence of the model parameters. Moreover, the validation loss remains consistently close to the training loss, suggesting good generalization across previously unseen local-environment configurations without significant overfitting.

The minimum validation loss achieved during training is $0.034$ at epoch $908$. The corresponding minimum validation mean absolute error is $0.026$ eV at epoch $978$, while the lowest validation root-mean-square error is $0.035$ eV at epoch $972$. The mean and median relative errors are $1.64\%$ and $1.10\%$, respectively. Collectively, these results demonstrate that the present deformation- and chemical environment aware neural network provides quantitatively accurate predictions of vacancy formation energies from the surrounding chemical environment, local atomic geometry, and local deformation descriptors.

\section*{Author contributions}
Tanvir Sohail:  Conceptualization,  Methodology,  Software, Formal analysis,  Investigation, Writing - Original Draft,  Writing - Review \& Editing,  Visualization; 
Swarnava Ghosh:  Conceptualization,  Methodology,  Software, Formal analysis,  Investigation,  Writing - Original Draft,  Writing - Review \& Editing,  Visualization,  Supervision,  Project administration.

\section*{Declaration of competing interest}

Declaration of competing interest
The authors declare that they have no known competing financial
interests or personal relationships that could have appeared to influence
the work reported in this paper.

\section*{Acknowledgement}
The authors would like to acknowledge Dr. David Rogers (ORNL) for critical reading and comments on an early draft of the manuscript. This research used resources of the Oak Ridge Leadership Computing Facility, which is a DOE Office of Science User Facility supported under Contract DE-AC05-00OR22725

During the preparation of this work the authors used OpenAI (GPT-5.6) in order to search literature, verify assumptions, correct grammar and revise language. After using this tool/service, the authors reviewed and edited the content as needed and takes full responsibility for the content of the published article.

\bibliographystyle{unsrt}
\bibliography{reference}
\end{document}